# Phase Distance Mapping: A Phase-based Cache Tuning Methodology for Embedded Systems

Tosiron Adegbija · Ann Gordon-Ross · Arslan Munir

**Abstract** Networked embedded systems typically leverage a collection of low-power embedded systems (nodes) to collaboratively execute applications spanning diverse application domains (e.g., video, image processing, communication, etc.) with diverse application requirements. The individual networked nodes must operate under stringent constraints (e.g., energy, memory, etc.) and should be specialized to meet varying applications' requirements in order to adhere to these constraints. Phase-based tuning specializes a system's tunable parameters to the varying runtime requirements of an application's different phases of execution to meet optimization goals. Since the design space for tunable systems can be very large, one of the major challenges in phase-based tuning is determining the best configuration for each phase without incurring significant tuning overhead (e.g., energy and/or performance) during design space exploration. In this paper, we propose phase distance mapping, which directly determines the best configuration for a phase, thereby eliminating design space exploration. Phase distance mapping applies the correlation between a known phase's characteristics and best configuration to determine a new phase's best configuration based on the new phase's characteristics. Experimental results verify that our phase distance mapping approach, when applied to cache tuning, determines cache configurations within 1% of the optimal configurations on average and yields an energy delay product savings of 27% on average.

## 1 Introduction and Motivation

Due to the pervasiveness of embedded systems, much research has focused on optimizations, such as improved performance and/or reduced energy consumption, to meet stringent design constraints imposed by physical size, battery capacity, cost, real-time deadlines, consumer market

T. Adegbija
Department of Electrical and Computer Engineering, University of Florida, USA E-mail: tosironkbd@ufl.edu

A. Gordon-Ross
Department of Electrical and Computer Engineering, University of Florida, USA E-mail: ann@ece.ufl.edu

A. Gordon-Ross
NSF Center for High-Performance Reconfigurable Computing (CHREC), University of Florida, Gainesville, USA

A. Munir
Department of Electrical and Computer Engineering, Rice University, USA
E-mail: arslan@rice.edu

competition, etc. However, system optimization is challenging due to numerous tunable parameters (e.g., cache size, associativity and line size [42], replacement policy [44], issue width [8], core voltage and frequency [40], etc.), many of which tradeoff design constraints, such as size versus performance, resulting in very large design spaces with many Pareto optimal systems. Thus, evaluating different designs in the design space either statically or dynamically at runtime to determine the best set of designs that result in optimal systems (i.e., design space exploration) can be very challenging. The advent of multicore systems further compounds optimization challenges due to a potential exponential increase in the design space when considering dynamic core dependencies and interactions [30], which change during runtime based on the currently co-scheduled tasks. Therefore, in order to meet these increasing challenges for future systems, optimization methodologies must be highly scalable to large design spaces and must be dynamic in nature.

Application-based tuning evaluates an application's characteristics and determines the best configuration (specific tunable parameter values) for the entire application's average execution requirements. However, since applications have varying/dynamic requirements during execution (i.e., different phases of execution) [17, 28, 37, 38], configurable/tunable hardware [11, 42, 44] enables dynamic adaptation to these requirements by specializing tunable parameters to the changing needs of the application. A phase is a length of execution where an application's characteristics, such as cache misses, instructions per cycle (IPC), branch mispredictions, etc., and therefore application requirements, remain relatively stable. To identify phases, the application's execution is broken into fixed or variable length intervals that are typically measured by the number of instructions executed. *Phase classification* [35, 37, 38] groups intervals with similar characteristics to form phases, using methods such as K-means clustering [22, 37], Markov predictors [38], etc. *Phase-based tuning* evaluates the application's characteristics and determines the best configuration for each phase of execution to best meet design constraints.

The interval length must be carefully defined in a phase-based tuning approach. Intervals that are too long tend to have less stable characteristics, thus making it difficult to determine the phase's best configuration. Intervals that are too short result in too frequent tuning, thus imposing significant accumulated tuning overhead in terms of energy and performance that may intrusively affect system operation/behavior. Since interval length selection is widely researched [12, 37, 38], and is thus not a focus of our work, we assume variable length intervals [12], which result in higher optimization potential [12].

A major challenge in phase-based tuning is determining the best configuration for each phase [17, 28] without incurring significant tuning overhead. Most previous methods [28, 42, 44] physically explore the design space by executing different configurations, recording the configurations' characteristics, and selecting the best configuration, however, this method incurs a large cumulative tuning overhead while executing inferior (non-optimal) configurations. To reduce tuning overhead, heuristics significantly prune the design space [13, 14, 30], however, since these heuristics still execute inferior configurations and incur tuning overhead. Analytical methods/models drastically reduce tuning overhead by directly determining/calculating/predicting the best configuration based on the design constraints and application characteristics [10, 15, 27, 28], however, most of these methods are either computationally complex (thus, adversely impacting performance and energy consumption) [10, 28] or not dynamic (i.e., not phase-based) [15, 27].

In this paper, we focus on reducing the computational complexity and tuning overhead of dynamic phase-based tuning by directly determining the phases' best configurations, with no design space exploration, using the correlations between a phase's characteristics and the phase's best configuration. We introduce *phase distance mapping* (PDM), which leverages these characteristic-to-configuration correlations to determine the best configurations for new phases based on the new phase's characteristics. PDM automatically analyzes applications, the applications' phases, and the phases' characteristics to determine the best configurations for the phases, thereby eliminating a priori designer effort while maintaining computational simplicity.

We define the *phase distance* as the difference between the characteristics of a characterized phase—a phase with a known best configuration—and an uncharacterized phase, and is used to estimate the uncharacterized phases' best configurations. PDM compares a single previously characterized phase—the *base phase*—with a new phase to determine the phase distance, and uses the phase distance to calculate the *configuration distance*, which is the difference between the tunable parameter values between two configurations. We also introduce *distance windows,* which define phase distance ranges and corresponding tunable parameter values. The distance window that the phase distance falls within (i.e., maps to) defines the tunable parameter values for the uncharacterized phase.

Using extensive analysis of sample phases from workloads representative of real-world embedded system applications (e.g., image processing, networking, etc.), we show that that given two phases, there is a strong correlation between the phases' phase and configuration distances. PDM uses the phase distance to calculate the configuration distance, and thus directly determines the uncharacterized phase's best configuration using *configuration estimation*. The configuration estimation results corroborates the correlation between the phase distance and configuration distance, revealing overall system energy delay product (EDP) savings of 24%, as compared to using a base/default configuration throughout execution, and configurations within 4% of the optimal.

However, our configuration estimation analysis and experiments revealed that phase correlation could be application-dependent (i.e., the phase correlation from one application may not be applicable to a different application). As a result of this non-uniformity and unpredictable phase correlation, accurately extracting phase correlation, if any, requires the designer to expend considerable design-time pre-analyzing effort on the applications and application phase characteristics to provide information for runtime configuration estimation decisions. To address these limitations, we also present DynaPDM, which uses computationally simple algorithms to facilitate and improve PDM's runtime phase correlation, eliminates the designer's a priori phase analysis, and produces more efficient results in terms of EDP savings, time to market, and is completely transparent to the designer.

We exemplify and evaluate PDM and DynaPDM using cache tuning for separate level one instruction and data caches, since caches constitute a large percentage of a microprocessor's energy budget and adapting cache configurations to application characteristics can significantly reduce the average memory access energy [14]. In order to optimize the energy without significantly adversely impacting the execution time, we use the overall system EDP as our evaluation metric (Section 5.1). Our evaluated cache tuning method determines the best cache configuration in terms of total size, line size, and associativity for reduced EDP [14, 17, 42]. Our experimental results using a variety of benchmarks indicate that using DynaPDM for cache tuning consistently improves EDP with respect to the base/default configuration. Results reveal that DynaPDM can determine configurations within 1% of the optimal configurations and achieves average EDP savings of 27%—an 8% improvement over PDM.

## 2   Background and Related Work

Phase distance mapping leverages fundamentals of phase classification, and simplifies phase-based tuning by directly determining a phase's best configuration without time consuming design space exploration. In this section, we describe work related to phase-based tuning, cache tuning, design space exploration, and phase classification.

### 2.1     Phase-based tuning and design space exploration

Much previous work focuses on tuning configurable hardware to the best configuration for a particular application for reduced energy consumption and/or improved performance. However, this tuning typically imposes tuning overheads in terms of design exploration tuning time, and energy and performance overheads while evaluating inferior configurations.

To reduce tuning time, several heuristic methods have been developed for searching the design space. Zhang et al. [42] proposed a configurable cache architecture that determined the Pareto optimal cache configurations trading off energy consumption and performance. The proposed heuristic searched the cache parameters in the parameters' order of impact on energy consumption, first determining the best cache size, followed by the best line size, and finally the best associativity. This method incurred tuning overhead by physically exploring the design space (i.e., the application executed in each configuration for a period of time to evaluate the configuration). L. Chen et al. [6] introduced a configuration management algorithm that searched the cache design space for the best configuration, which leveraged Zhang et al.'s [42] energy-impact parameter search ordering, and incurred similar tuning overheads.

To reduce tuning overhead, several methods eliminated design space exploration, thus eliminating any tuning overhead due to executing inferior configurations. Gordon-Ross et al. [15] proposed a one-shot approach to cache configuration using a cache tuner that non-intrusively predicted the best cache configuration using an oracle-based approach [20]. This method monitored an application's memory access pattern and analytically predicted the best cache configuration based on these patterns. However, the oracle hardware introduced significant tuning overhead when active. Ghosh et al. [10] proposed an analytical model to directly determine the cache configuration based on the designer's performance constraints and application characteristics, however, the model's computational complexity incurred energy and performance overheads. Even though these methods reduced the tuning overhead, these methods were not phase-based.

Phase-based tuning, as opposed to application-based tuning, evaluates an application's characteristics and determines the best configuration that satisfies design constraints for each application phase. To adhere to an application's changing execution requirements, Hajimir et al. [17] used a cache model for phase-based tuning that used changes in application characteristics to determine when to change the cache configuration and presented a dynamic programming-based algorithm to find the optimal cache configuration. The cache model evaluated the energy consumption and performance for every possible cache configuration for each phase and selected the lowest-energy configuration for the different phases during runtime. Albonesi et al. [1] presented a method that adaptively changed the cache associativity by analyzing the application's software at compile time or using dynamic profiling to determine the application's associativity requirements. This method disabled cache ways during periods where full cache functionality was not required while limiting the performance degradation to within an allowable threshold based on design constraints. Gulati et al. [16] proposed a scheduling scheme that exploited varying application characteristics by using an efficiency threshold for dynamic task-to-core allocation in flexible-core chip multiprocessors, wherein this flexibility enabled small cores to be aggregated to form larger logical cores. The proposed scheme focused on improving throughput by scheduling tasks to processors based on how efficiently the processor executed the tasks—tasks that achieved higher efficiency than the threshold were given higher priority in terms of number of cores allocated to that task. Peng et al. [28] proposed a phase-based tuning algorithm that managed a configurable cache on a per-phase basis and attempted to reduce performance loss due to unnecessary reconfigurations. The algorithm monitored cache performance (i.e., cache miss rates) during execution and modified the configuration based on the observed performance.

Chaver et al. [4] presented a phase-based adaptive instruction fetch mechanism that used an offline profiling step to statically divide applications into phases, and determined system resource requirements (e.g., trace cache size, branch target buffer size, etc.) based on the phases' characteristics. Gordon-Ross et al. [11] investigated the benefits of phase-based tuning over application-based tuning with respect to energy consumption and performance, and showed that the tuning overhead due to cache flushing and write backs was minimal. Results showed that phase-based tuning yielded improvements of up to 37% in performance and 20% in energy over application-based tuning, however, to maximize phase-based tuning savings, phase changes must be quickly detected and phases accurately characterized/classified.

### 2.2    Phase classification

Phase classification can be done dynamically at runtime (online) or statically (offline) and is a widely studied research area. Much research has substantiated that dynamically leveraging phase characteristics reveals a finer grained optimization potential by specializing the configurations to the different phases of execution. Sherwood et al. [34] studied the time varying behavior of applications using the SPEC 95 benchmarks and showed that applications have periodic patterns and phase-based characteristics with respect to several hardware metrics (e.g., cache size, branch prediction, value prediction, IPC, etc.). The authors observed that the metrics with the largest impact on energy consumption and performance tended to change simultaneously, thus denoting phase change occurrences (i.e., when the application transitions from one phase to another).

In order to detect these periodic patterns and determine the patterns' durations, Sherwood et al. [35] proposed using basic block distribution analysis as an automated approach for finding the periodic and phase-based characteristics of applications for phase classification. The basic block distribution represents the entire application's behavior and can be obtained using a basic block profiler, which measures the number of times each basic block is executed, thus obviating the need for cycle accurate simulations. The authors also showed that basic block distribution analysis is highly correlated with architecture-dependent application characteristics (e.g., cache miss rates, branch miss rates, etc.). Due to prohibitively long cycle-accurate simulation times in computer architecture research, a small set of phases that provide an accurate and efficient representation of an application's execution need to be identified. To identify these phases, the authors created SimPoint [36]. SimPoint used machine learning techniques to identify the application's phases by analyzing basic block vectors that contained the frequency of executed code and used clustering algorithms to choose the phases that represented the application's complete execution. In [37], the authors further showed that phase characteristics could be collected using basic block vector profiles for offline classification or through dynamic branch profiling for online classification, which provides more accurate phase classification.

For generalized applicability, these phase classification methods used basic block vectors, which are architecture-independent, to classify phases. Later research showed that architecture-dependent characteristics could also effectively classify phases. Balasubramonian et al. [2] used cache miss rates, cycles per instruction (CPI), and branch frequency characteristics to detect changes in application characteristics for cache tuning, and found that these characteristics were effective for phase classification. Shen et al. [33] showed that data locality was well suited for phase classification by using a method that combined data locality profiling and runtime prediction to predict recurring application phases.. Dhodapkar et al. [7] proposed a method to determine phase changes by examining the application's working set (i.e., address access locality). The authors found a relationship between phases and an interval's working set, and concluded that phase changes could be detected by detecting changes in the working set.

## 3    Key Terminology and Architectures

In this section, we present key terminology used in describing PDM, and the major architectural components needed for implementing PDM, including the configurable cache architecture and the phase tuning architecture.

### 3.1    Key terminology

The *phase distance* is the difference between the characteristics of a characterized phase—phase with a known best configuration—and an uncharacterized phase and is used to estimate the uncharacterized phase's best configuration. PDM compared a single previously characterized phase—the *base phase*—with a new phase to determine the phase distance. For example, for cache tuning, PDM used the instruction and data cache miss rates to characterize the phases and normalized the uncharacterized phases' cache miss rates to the base phase's cache miss rates to determine the phase distance. PDM used the phase distance to calculate the *configuration*

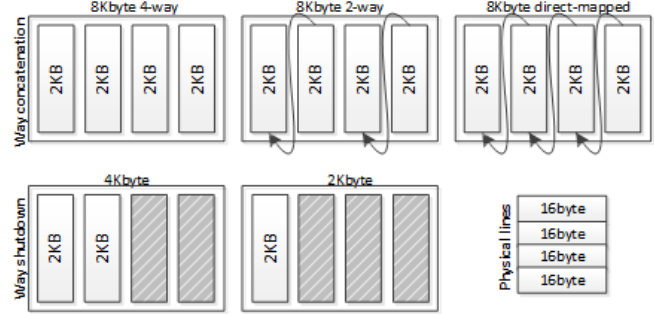

**Fig. 1** Configurable cache architecture

*distance*—the difference between the tunable parameter values of two configurations. Given two phases *I* and *J*, *I*'s best configuration may be the base phase's associativity increased by a power of two, while *J*'s best configuration may be the base phase's associativity reduced by a power of two. These parameter value changes represent *I* and *J*'s configuration distances from the base phase. Finally, *distance windows* define phase distance ranges and corresponding tunable parameter values. Each distance window that the phase distance falls within (i.e., maps to) contains a configuration distance from the base phase's best configuration, and this configuration distance defines the tunable parameter values (i.e., best configuration) for the uncharacterized phase.

### 3.2  Configurable cache architecture and design space

Prior work has developed various configurable cache architectures and dynamic tuning methods to search the configuration design space, which consists of all the different configurations/combinations of the tunable parameter values. Motorola's M*CORE processor [24] provided per-way configuration using way management, which allowed ways to be shut down or designated as instruction only, data only, or unified. Modarressi et al. [26] developed a cache architecture that was partitioned and resized dynamically to improve the performance of object-oriented embedded systems.

Our memory hierarchy consists of configurable, private, separate level one (L1) instruction and data caches. Typical modern day microprocessors also usually include a shared L2 cache; however, since we only tune the L1 cache in this study, we ignore the L2 cache in our discussions. The configurable caches are based on Zhang et al.'s [42] highly configurable cache, which provides runtime-configurable total size, associativity, and line size using a small bit-width configuration register. Zhang's configurable cache does not increase cache access time since the cache imposes no overhead to the critical path. The configurable cache has served as the basis for several newer architectures [12, 14, 41] and can be easily extended to state of the art, more complex architectures, such as heterogeneous multicore systems [30].

To evaluate phase distance mapping, we define a *base cache configuration* for comparison purposes. The base cache configuration, which is an average configuration representing typical embedded microprocessors [42] that might execute our experimental applications (Section 5), is an 8 Kbyte cache composed of four configurable banks, each of which can operate as a separate way (i.e., the base cache is a 4-way set associative cache), and a logical line size of 64 bytes. Figure 1 depicts the configurable cache architecture. The configuration register provides configurable associativity by logically concatenating the ways, offering an 8 Kbyte direct-mapped or 2-way set associative cache, and/or shutting down ways, offering a 4 Kbyte direct-mapped or 2-way set associative cache or a 2 Kbyte direct-mapped cache. All cache sizes offer a configurable line size of 16, 32, or 64 bytes by using a base, physical line size of 16 bytes and fetching additional physical cache lines for larger, logical line sizes. Due to the bank layout for way shut down, 2 Kbyte 2-way or 4-way set associative and 4 Kbyte 4-way set associative caches are not feasible using this configurable hardware, thus both the instruction and data caches each have eighteen possible cache configurations, resulting in a large design space that necessitates an efficient method for determining the best configurations for dynamic tuning.

**3.3**   Phase tuning architecture

Figure 2 depicts our phase tuning architecture for a sample dual-core system, which can be extended to any *n*-core system. On-chip components include the processing cores that are connected to private, separate L1 instruction and data caches and the phase characterization hardware. Without loss of generalization, the level one caches are directly connected to off-chip main memory, and since this hierarchy implies that there is no dependence between the caches, the caches can be tuned independently. Phase characterization hardware includes a *tuner*, a *phase classification module* that classifies an application's phases, a *PDM module*, which includes a *distance window table*, which stores the distance windows and serves as a lookup table for the configuration distances when phases are characterized, and a *phase history table*.

The tuner orchestrates the phase characterization process (Section 4.1), which includes implementing PDM, by executing the phase in each potential configuration for one tuning interval, gathering cache statistics, and calculating the EDP. The PDM module implements the algorithms presented in this work (Section 4.2) for determining a phase's best configuration. After PDM determines a phase's best configuration, the phase is designated as a *characterized phase* and is added to the phase history table, along with the phase's best configuration. We note that in the case of our studied cache hierarchy, the best configuration stored in the *phase history table* represents both the best instruction and data cache configurations. The distance window table's structure is similar to the phase history table's structure [37], and can be easily implemented as a software- or hardware-based lookup table. The number of distinct phases and distance windows dictates the number of entries in the phase history and distance window tables, respectively. The maximum number of distance window entries is constrained by the total number of distinct phases across all applications running on the system, thus, the distance window table's size should be less than or equal to the phase history table's size. To minimize the hardware or memory overhead from these tables, the number of entries can be constrained, and the least recently used entries can be evicted when necessary. However, the actual table sizes are determined by the design constraints of the embedded system.

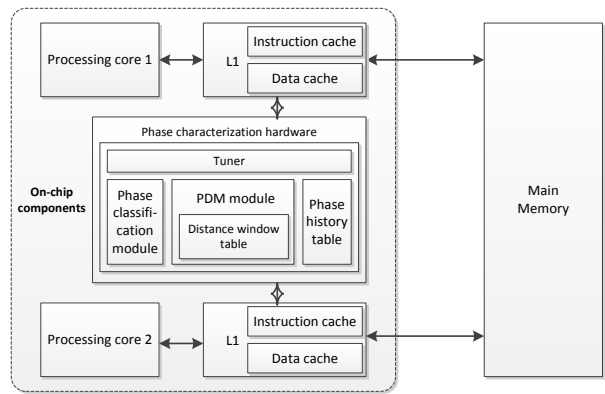

**Fig. 2** Phase tuning architecture for a sample dual-core system

Prior research using similar table structures showed that these structures have very little or no effect on overall system area, performance, and/or energy consumption [37, 38], and the work proposed herein to incorporate phase distance mapping will not significantly increase/impact these overheads.

## 4  Phase Distance Mapping

Phase distance mapping reduces tuning overhead by directly determining a phase's best configuration by evaluating the correlation between the phase distance and the configuration distance. In this section, we elaborate on how this correlation is leveraged to determine a phase's best configuration and present our algorithm for configuration estimation using phase distance mapping. Even though we exemplify phase distance mapping using cache tuning, we generalize our discussions for any tunable hardware and include cache tuning specifics when necessary.

### 4.1    PDM Overview

Phase classification groups intervals that show similar characteristics into phases such that a phase's characteristics are relatively stable during the phase's execution. As a result of this relative stability, the same configuration can be used for the phase's duration. Therefore, our foundation for phase distance mapping is the hypothesis that the more disparate two phases' characteristics are, the more disparate the phases' best configurations are likely to be, enabling the mapping of the distance between phases to the distance between the best configurations.

We calculate the phase distance based on the phase space, which is the set of all of an application's distinct phases. Since phase classification is not the focus of this study, we assume that phase classification has already been applied to the application (using any arbitrary method, such as offline phase classification [35] or online runtime phase tracking and prediction [38]), which produces the application's different phases and the phases' characteristics. Since we study cache tuning and previous work showed that cache miss rates can accurately determine a phase's characteristics [30, 37], we classify the different phases using the phases' cache miss rates. Since comparative cache evaluation is most effective when the caches have the same configuration, we gathered the phases' cache miss rates for the base cache configuration (Section 3).

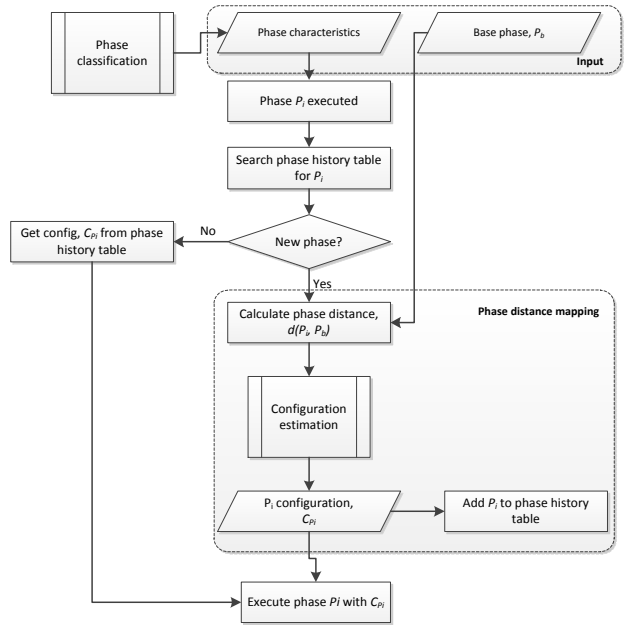

**Fig. 3** Phase characterization using configuration estimation

Figure 3 illustrates phase characterization using configuration estimation, which takes as input the classified phases and the phases' characteristics, which are output from phase classification. One phase is designated as the *base phase* $P_b$. The base phase is the phase to which subsequent phases are compared to calculate the phase distance, thus, to maximize EDP savings, the base phase should reflect the systems prominent application domain (e.g., image processing, networking). For a small, application-domain specialized system with a small phase space, designating the base phase can be easily done manually at design time, however, this method is infeasible for large, general-purpose systems with large phase spaces. For large systems, designers can use cluster analysis (e.g., k-means clustering [21], graph-based models [43]) to partition the phase space into different domains, and a phase that most closely represents the largest cluster (most prominent domain) is designated as the base phase.

In order to designate and characterize the base phase at design time, the designer requires a priori knowledge of the system's intended application domain(s), and the design space must be small enough or the designer must have an efficient design exploration method to afford quick design-time tuning. After designating the base phase, the designer can then use any tuning method (e.g., [42]) to determine the base phase's best configuration.

However, for general-purpose systems, where the application domain(s) are not known a priori, to maximize EDP savings, the base phase should be dynamically designated at runtime. Using a dynamic base phase requires the phase classification module to cluster executing phases by application domain, monitor the domains' numbers of phases, designate a base phase from the prominent domain, and re-designate new base phases when the prominent domain changes. The prominent domain is the application domain with the largest number of phases. When a new phase executes, this phase is added to the appropriate cluster of phases belonging to the same application domain as the executing phase (e.g., networking, image processing, etc.). If no phases have been previously executed that belong to the same domain as the executing phase, a new cluster is formed for that application domain. A counter tracks the number of phases in each application domain, and the domain with the largest number of phases is designated as the prominent domain, from which the base phase is then arbitrarily selected.

When a phase $P_i$ is executed, the first step in phase characterization is to search the *phase history table* for $P_i$. If $P_i$ is in the *phase history table*, $P_i$ has already been executed and the best configuration $Config(P_i)$ has already been determined. The hardware is configured to $Config(P_i)$ and phase $P_i$ executes in $Config(P_i)$. If $P_i$ is not in the *phase history table*, $P_i$ is a new phase and the difference between $P_i$'s characteristics and the base phase's characteristics $d(P_b, P_i)$ (i.e., the phase distance) is calculated.

The phase distance can be calculated using either a single phase characteristic or multiple phase characteristics. In this work, we use a single phase characteristic, the cache miss rate, to calculate $d(P_b, P_i)$, by normalizing $P_i$'s instruction and data cache miss rates to $P_b$'s instruction and data cache miss rates. This normalization enables quick comparisons of disparate configurations' miss rates. This single-characteristic method is suitable for tuning single components, such as private instruction and data caches that do not have dependencies. In systems with multiple tunable hardware or tunable component dependencies, a multi-characteristic method, such as one that evaluates the cache miss rates and IPC, provides a more holistic view of the phase characteristics and is the focus of our future work.

After the phase distance is calculated, the phase distance is used as input to configuration estimation.

**4.2    Configuration estimation**

As an initial step to ascertaining the correlation between the phase distance and configuration distance, we statically evaluated the cache characteristics of a set of applications, and studied the correlation between the phase distances and the best configurations for each phase (determined by exhaustive search). Using the results of our studies, we developed a *configuration estimation* algorithm that leverages the correlation between the phase distance and configuration distance to determine the best configuration for a phase.

We empirically developed and refined the configuration estimation algorithm by studying the impact that the different configurations have on the phases' characteristics. Since most embedded systems run single applications or a set of applications within the same domain, configuration estimation can be application domain-specialized with respect to the underlying tunable hardware. However, we point out that even though our configuration estimation is domain-specialized, the algorithm is generalized and can be easily adapted to different domains and tunable hardware. We generalized our configuration estimation algorithm to a variety of common embedded systems application domains, such as networking, image processing, cryptography, and data compression. However, since the majority of our studied applications involved image rotation (application details are presented in Section 5), we specialized the configuration estimation algorithm to an image processing domain by using a base phase from an image rotation application.

Configuration estimation leverages the underlying tunable hardware by considering the impact of the different parameter values on the energy consumption and performance [18, 41, 42]. For example, direct-mapped caches consume less power per access than 4-way set associative caches since only one data array and one tag are read per access, rather than four data arrays and four tags. However, direct-mapped caches can have higher cache miss rates than set associative caches, resulting in more total energy consumption when considering the miss penalties in terms of stall time and power to access the next memory level(s). Even though increasing the cache associativity increases the power per access, the cache miss rate may decrease enough to result in an overall decrease in energy consumption. However, this concept suffers from diminishing returns as increasing the reduction in miss rate (i.e., increasing the set associativity) will eventually not outweigh the increase in power per access. Since this well-known trend is not isolated to cache parameters, configuration estimation must consider diminishing returns for all tunable parameters with similar trends. Our configuration estimation algorithm considers diminishing returns using threshold values for each tunable parameter. A threshold value is the specific parameter value at which further increases in the parameter value may result in increased energy consumption or reduced performance.

**Algorithm 1** Configuration estimation

**Inputs:** $C_b$, $A_b$, $L_b$, $C_{min}$, $C_{max}$, $A_{min}$, $A_{max}$, $L_{min}$, $L_{max}$, $C_{THR}$, $A_{THR}$, $L_{THR}$, $R_1$, $R_2$, $R_3$, $R_4$, $R_5$, $R_6$, $R_7$, $D = d(P_b, P_i)$
**Outputs:** $C_i$, $A_i$, $L_i$  //output best cache size, associativity, and line size

```
 1   Ci ←Cb, Ai ←Ab, Li ←Lb   //initialize Ci, Ai, and Li
 2   //determine which distance window the phase distance maps to \
 3   //and determine the best configuration
 4   Switch (D)
 5      case R1, R2, R7:
 6         Ci ←CTHR
 7      break
 8      case R3:
 9         if Cb == Cmin then
10            Ci ←Cb * 2
11         else
12            Ci ←CTHR
13         end
14         if Ab = Amin then
15            Ai ←Ab * 2
16         end
17      break
18      case  R4:
19         Ci ←CTHR
20         if Ab != Amax then
21            Ai ←Ab * 2
22         end
23         if Lb != Lmin then
24            Li ←Lb/2
25         end
26      break
27      case R5:
28         Ci ←CTHR
29         if Ab = 1 then
30            Ai ←ATHR
31         end
32      break
33      case R6:
34         if Cb != Cmax then
35            Ci ←Cmax/2
36         end
37      break
38   end
```

Algorithm 1 depicts the configuration estimation algorithm, which defines the correlation between the phase distance and configuration distance. The algorithm's inputs are: the base phase's best configuration in terms of cache size $C_b$, associativity $A_b$, and line size $L_b$; the configurable cache's minimum and maximum sizes $C_{min}$ and $C_{max}$, associativities $A_{min}$ and $A_{max}$, and line sizes $L_{min}$ and $L_{max}$, respectively; size, associativity, and line size threshold values $C_{THR}$, $A_{THR}$, and $L_{THR}$, respectively; distance windows $R_1$ through $R_7$; and the phase distance $D$. The algorithm outputs phase $P_i$'s determined best cache size, $C_i$, associativity $A_i$, and line size $L_i$.

We empirically determined the threshold cache size, associativity, and line size values as 8 Kbyte, 2-way, and 64 byte, respectively. For example, Figure 4 illustrates how we determine the associativity threshold value in terms of EDP (Joule seconds) for three image rotation phases from our studied applications (Section 5 details the EDP calculation and application phases). In these results, the instruction cache configuration is arbitrarily fixed at the base configuration and the data cache associativity is varied while holding the data cache's size and line size fixed at the base configurations. Since increasing the associativity from 1-way to 2-way results in a decrease in EDP and further increasing the associativity to 4-way results in an increase in EDP, the associativity threshold value is 2. We similarly determined the size and line size threshold values. Even though this is an expected result for a simple trend, this empirical analysis can be used for any tunable parameter with any number of parameter values. Even though the threshold values can be generalized for any application domain, the specific threshold values will vary across different application domains due to different cache locality behavior. Therefore, for configuration estimation to be most effective, the threshold values should be application domain-specialized. We note, however, that since our experiments considered phases from diverse application domains, we used generalized threshold values, which underestimate the effectiveness of our configuration estimation algorithm.

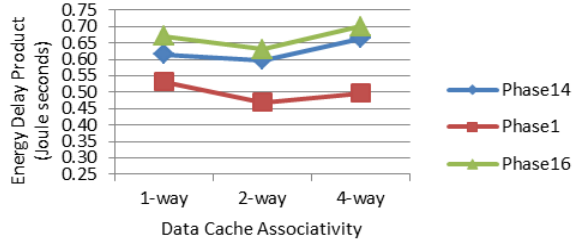

**Fig. 4** Associativity threshold value determination using diminishing return effects on the energy delay product for varying data cache associativities.

Distance windows are phase distance ranges that represent an uncharacterized phase $P_i$'s configuration distance from the base phase $P_b$ when changing a parameter's value to another value (e.g., increasing the associativity: $A_b * 2$). Each distance window has a maximum $Win_M$ and minimum $Win_L$ and a phase distance $D$ maps to the distance window in which $D$ is bounded by (i.e., $Win_L \leq D \leq Win_M$). For our experiments, we created distance windows using a base phase from an image rotation application and evaluated how the parameter values changed for the different phases' optimal configurations (determined by an exhaustive search) with respect to the base phase's configuration. The distance windows relate directly to all of the characteristics used to evaluate $D$ and are applicable to all the tunable parameters represented by $D$. For example, since we use the cache miss rate to evaluate $D$, the distance window bounds relate directly to the actual cache miss rate values and are applicable to all of the tunable parameters (cache size, associativity, and line size). We determined that the seven distance windows: $R_1 = [0,0.25]$, $R_2 = (0.25,0.5]$, $R_3 = (0.5,0.75]$, $R_4 = (0.75,1.25]$, $R_5 = (1.25,1.5]$, $R_6 = (1.5,2.5]$, and $R_7 = (2.5,\infty]$), sufficiently cover all the phase distances between the base phase and all of the other phases. The distance windows' bounds represent the normalized difference between $P_i$'s and $P_b$'s cache miss rate. The phase distance $D$ maps to these distance windows such that if $0 \leq D < 0.25$, $D$ maps to $R_1$, if $0.25 \leq D < 0.5$, $D$ maps to $R_2$, etc. In general, the number of distance windows can vary based on a system's intended applications and the applications' phases, the distance windows are specialized based on the evaluated characteristic (e.g., cache miss rates or IPC), and if a multi-characteristic method is used for evaluating $D$, only one set of distance windows is necessary to represent all of the tunable parameters.

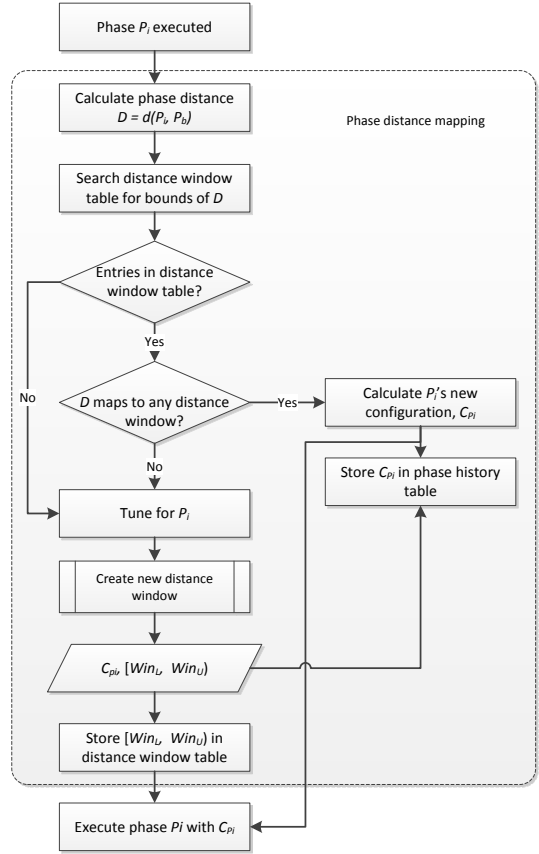

**Fig. 5** Phase characterization using dynamic phase distance mapping.

For each phase $P_i$, the configuration estimation algorithm is executed twice, once for the instruction cache and once for the data cache. First, the algorithm assigns initial values to $C_i$, $A_i$, and $L_i$ as $C_b$, $A_b$, and $L_b$, respectively (line 1), which represent default values for $C_i$, $A_i$, and $L_i$. Default values are used because some configuration distances in some distance windows require no parameter value change for some parameters. Next, the algorithm determines which distance window the phase distance $D$ maps to (line 4) and determines $P_i$'s best configuration based on the configuration distance for the corresponding distance window. If a distance window does not specify a change to a parameter value, then $C_i$, $A_i$, and $L_i$ remain as the default values. For example, if phase $P_2$ is the next phase to be executed and $D = 1.08$, the algorithm determines that $D$ maps to distance window $R_4$ (line 18), and determines $C_i$, $A_i$, and $L_i$, based on the configuration distance for $R_4$ (lines 19 – 25).

We note that even though PDM showed good average EDP savings using configuration estimation (results are detailed in Section 5), PDM had several limitations. First, the designer was required to statically define the distance windows based on the anticipated applications, which limits the configuration estimation's applicability to dynamic systems where applications are not known a priori. PDM using configuration estimation also required the designer to designate the base phase such that the base phase represented the system's prominent application domain. Appropriate base phase designation was critical since the EDP savings were strongly affected by how well the base phase represented the entire system. Thus, we created DynaPDM, which refined PDM to address these limitations by dynamically analyzing the applications, applications' phases,

**Algorithm 2** Dynamic distance window creation

```
Input: S_d, D, WinU_max
Output: Win_L, Win_U   //output new distance window's lower and upper bound

  1   if D < S_d then    //create the first distance window
  2      Win_L = 0
  3      Win_U = S_d
  4   else if D > WinU_max then    //create the last distance window
  5      Win_L = WinU_max
  6      Win_U = ∞
  7   else    //create distance windows below maximum upper bound
  8      Win_L = x | (x ≤ D, x mod S_d = 0, x + S_d > D)
  9      Win_U = x + S_d
 10   end
```
_________________________________________________________________

and configurations, thereby eliminating designer effort while maintaining the computational simplicity, low tuning overhead, and phase-based nature of PDM.

### 4.3   Dynamic phase distance mapping

DynaPDM dynamically creates and stores distance windows in the distance window table as phases execute. Figure 5 overviews the DynaPDM flow. When a new phase $P_i$ is executed (i.e., $P_i$ is not in the phase history table) and $P_i$'s phase distance $D$ maps to an existing distance window, $P_i$'s new configuration $config(P_i)$ is calculated, stored in the phase history table, and the system is configured to $config(P_i)$. If $D$ does not map to any distance windows or the distance window table is empty (special case at system startup), a new distance window is created.

Algorithm 2 dynamically creates a new distance window during runtime and takes as input: the distance window size $S_d$, $D$, and the maximum upper bound for the distance window $WinU_{max}$. The length of each distance window (i.e., the difference between $Win_U$ and $Win_L$) is determined by the distance window size $S_d$. $S_d$ directly affects the size of the distance window table since a larger distance window size indicates fewer distance windows (i.e., fewer distance window entries), while a smaller distance window size indicates more distance windows. To make the distance window sizes amenable to runtime changes, $S_d$ defaults to 0.25 at system start up and dynamically changes during execution. We empirically define $S_d = 0.25$ as the minimum distance window size since smaller sizes would drastically increase the number of distance windows without improving the tuning efficiency, and because similar phases could map to different distance windows, resulting in inaccurate characterization of these phases. As phases execute, if two contiguous distance windows have the same configuration distance, these two windows are combined into a single distance window and the distance window size is increased by 0.25. Alternatively, $S_d$ may be statically defined and maintained throughout execution. We empirically determined $S_d = 0.5$ as a generally suitable static value based on a variety of training applications representative of common embedded processor applications (detailed in Section 5). However, statically defining $S_d$ offers no clear advantages over dynamically defining $S_d$.

Creating either static or dynamic distance windows follow the same procedure. If $D < S_d$, the algorithm sets $Win_L$ to 0 and sets $Win_U$ to $S_d$ (lines 1 − 3). $WinU_{max}$ is optional and represents the maximum number of new distance windows $D$, such that if $D > WinU_{max}$, $D$ maps to $WinU_{max} < D < \infty$ (lines 4 − 6). $WinU_{max}$ defaults to infinity, which may improve the configurations' efficacies using unlimited smaller, thus more accurate, finer-grained distance windows, but could exhaust hardware resources. Defining $WinU_{max}$ restricts the number of distance windows to $WinU_{max}/S_d$ and may reduce accuracy since all phases with $D > WinU_{max}$ map to the same distance window, which may not accurately define those phases' configuration distances. If $S_d < D < WinU_{max}$, the next

**Algorithm 3** Initializing distance windows

```
Input: n, config(P_b)
Output: config(P_i)

    1   if n == 1 then    //set initial configuration
    2       config(P_i)_init ← config(P_b)
    3   else    //set most similar phase's best configuration as initial
    4       for j ←1 to n do
    5           D_j ← d(P_i, P_j)
    6       end
    7       P_msp ← P_j | D = min(D_j)
    8       config(P_i)_init ← config(P_msp)
    9   end
   10   //adjust configurations while EDP improves
   11   config(P_i) ← adjust(config(P_i)_init)
   12   if EDP[config(P_i)] ≤ EDP[config(P_i)_init] then
   13       config(P_i)_init ← config(P_i)
   14       goto line 11
   15   end
   16   updateDistanceWindowTable()   //store new configuration distance
   17   return config(P_i)
```

value smaller than $D$ and divisible by $S_d$ is selected as $Win_L$ for that distance window and $Win_U$ is set as $Win_L + S_d$ (lines 7 – 9).

Since there is no configuration distance information at system startup, the first executed phase, designated as the base phase, is tuned using any efficient tuning method (e.g., [41]) to determine that phase's best configuration. Our experimental results (Section 5) showed that the choice of the base phase does not affect the tuning efficiency since the distance windows are dynamically created at runtime to adapt to the executing applications. We reiterate that DynaPDM only requires a single arbitrary base phase, which alleviates the effort required in designating a base phase that represents the prominent application domain.

For a new executing phase $P_i$ that does not map to any distance window, DynaPDM determines $P_i$'s most similar phase $P_{msp}$ and uses $P_{msp}$'s best configuration as $P_i$'s initial configuration. $P_{msp}$ is the phase with the minimum phase distance $D$ from $P_i$ among all the previously executed phases. Using $P_{msp}$'s best configuration as $P_i$'s initial configuration represents a configuration that is presumably closer to $P_i$'s best configuration thus exploiting any potential phase correlation. DynaPDM then adjusts $P_i$'s configuration to determine $P_i$'s best configuration without significant tuning overhead, and this best configuration is used to initialize $P_i$'s distance window. To adjust $P_i$'s configuration, DynaPDM gradually increases the cache size, associativity, and line size, individually, while holding the other parameters fixed. While adjusting the configuration, DynaPDM consistently monitors the executing configuration's EDP (calculated by the tuner) and stops adjusting the configuration when an executing configuration achieves no EDP savings over previously explored configurations. The configuration with the lowest EDP is then designated as $P_i$'s best configuration. Our experiments (details in Section 5) reveal that DynaPDM can determine the best configuration after exploring as few as three configurations. After determining $P_i$'s best configuration, DynaPDM uses this configuration to initialize $P_i$'s distance window.

When a new executing phase maps to an existent distance window, DynaPDM directly determines the phase's configuration using that distance window's configuration distance. During the first execution, DynaPDM adjusts the phase's configuration in order to achieve a configuration closer to the phase's best configuration in the case of an inaccurate configuration distance. A configuration distance may be inaccurate if the distance window was initialized using an inferior, non-optimal configuration. Also, adjusting the phase's configuration helps DynaPDM determine a configuration closer to the best configuration in the rare case where a phase maps to an existent distance window, but requires a different configuration than the configuration determined by configuration distance in that distance window. However, our experimental results (Section 5.2) showed that even in these cases, DynaPDM significantly improved the EDP over using the base configuration, and adjusting the phase's configuration only further increased the EDP savings. After determining the new phase's best configuration, the distance window is updated, if necessary.

**Table 1** Experimental workloads

| Domain | Workload |
|---|---|
| Image processing | rotate-16x4Ms32w8 |
| | rotate-16x4Ms4w8 |
| | 64M-rotatew2 |
| | rotate-4Ms4w1 |
| | rotate-520k-270deg |
| | rotate-color-4M-90degw1 |
| Networking | 4M-check |
| | 4M-reassembly |
| | 4M-tcp-mixed |
| | ippktcheck-8x4M-4Worker |
| | ipres-6M4worker |
| MD5 checksum | md5-128M4worker |
| | md5-32M4worker |
| | md5-4M |
| Empty | empty-wld |
| Code compression | huffde-all |
| Video | x264-4M |
| | x264-4Mq |
| | x264-4Mqw1 |

In order to maintain the consistency of EDP savings achieved by DynaPDM's determined configurations when executing persistent phases (i.e., phases that reoccur several times throughout the system's lifetime) DynaPDM periodically monitors the EDP after executing a previously characterized phase and compares the current EDP to the previously monitored EDP for the same phase. Since most modern microprocessors contain performance monitoring units, periodically monitoring the EDP will not constitute any significant additional overhead. If the phase's current execution results in a significant EDP increase compared to previous executions, DynaPDM determines a new best configuration for that phase, using the previously determined configuration as the initial configuration. The phase's new best configuration is then stored in the phase history table. A significant EDP increase, instigating the need for DynaPDM to determine a new best configuration for a previously characterized phase, may result from a change in the hardware behavior due to changes in the input stimuli or external factors (e.g., temperature).

Algorithm 3 initializes the distance windows and updates the phase history and distance window tables. The algorithm takes as input the number of previously executed phases $n$ and the base phase's best configuration, and outputs the executing phase $P_i$'s best configuration $config(P_i)$. Since the algorithm's optimization goal is to determine a configuration for each phase with an EDP less than or equal to the base configuration's EDP $EDP[config(P_i)_{base}]$, all new phases default to the base configuration as the best configuration. The base configuration is initially stored in the phase history and distance window tables as the lowest EDP configuration. The algorithm monitors the EDP after every tuning interval, and only updates the phase history and distance window tables when the current EDP is less than the stored EDP. DynaPDM executes $P_i$ in each explored configuration for a tuning interval of 500,000 cycles, which is long enough to warm up the cache and for miss/hit rates to stabilize. If $P_i$ completes execution in fewer cycles than required for DynaPDM to determine the best configuration, $P_i$ begins subsequent executions in the stored lowest EDP configuration and DynaPDM continues exploring the configurations to determine $P_i$'s best configuration.

The algorithm uses the base phase $P_b$'s configuration $config(P_b)$ for $P_i$'s initial configuration $config(P_i)_{init}$ (lines 1 – 2) if only $P_b$ has been previously executed ($n = 1$). Otherwise, the algorithm uses the most similar phase $P_{msp}$'s best configuration $config(P_{msp})$ as $P_i$'s initial configuration (lines 3 – 9). The algorithm then adjusts $P_i$'s configuration while new configurations achieve lower EDP than previously explored configurations (lines 11 – 15) and stores the configuration with the lowest EDP as $P_i$'s best configuration. The algorithm then updates $P_i$'s distance window table with the configuration distance from the base phase and uses $config(P_i)$ for $P_i$'s subsequent executions.

## 5 Experimental Results

We evaluate PDM and DynaPDM by comparing a system that switches to the best configurations, as determined by phase distance mapping, for each phase to a system fixed with the base cache configuration. We present our experimental setup and the results obtained from both PDM using configuration estimation and DynaPDM, and compare the results obtained by DynaPDM to those obtained by PDM.

### 5.1 Experimental Setup

We selected nineteen workloads from the EEMBC Multibench benchmark suite [9], which is an extensive suite of multicore benchmarks that primarily target the embedded market and model a wide variety of realistic applications. Table 1 depicts the workloads used in our experiments. Each Multibench workload is a collection of kernels working on a specific dataset. Our selected workloads covered diverse processing tasks, such as image rotation for different colors/sizes, internet protocol (IP) packet checking, IP packet reassembly, transmission control protocol (TCP) processing, video encoding, md5 message-digest algorithm checksum calculation, Huffman decoding, etc. Since each workload was a collection of specific compute kernels, each of which performed a single task or a combination of similar tasks, the kernels essentially represented a single phase of execution. Therefore, without loss of generality, we assumed that each workload represented a different phase.

We simulated the system using Perl scripts for each phase to completion for the optimal, base, PDM, and DynaPDM configurations for all executions of each phase. To gather cache miss rates, we used GEM5 [3] to model a homogeneous dual core system with separate, private L1 instruction and data caches. We used McPAT [23] to calculate the system's total power

**Table 2** Core microarchitectural parameters

| Architectural Configuration | |
|---|---|
| Processing Cores | 2 |
| Clock Rate | 2 GHz |
| Functional Units | 2 IntAlu, 1 FPAlu, 1 Mult/DivAlu |
| Issue Width | 1 |
| Physical Registers | 32 Int, 32 FP |
| L1 Instruction and Data Caches | |
| Cache size | 2 Kbyte – 8 Kbyte |
| Associativity | 1-way – 4-way |
| Line size | 16 byte – 64 byte |

consumption and evaluate the system's energy efficiency using the EDP in Joule seconds:

$$EDP = system\_power * phase\_running\_time^2$$
$$= system\_power * (total\_phase\_cycles/system\_frequency)^2$$

where *system_power* includes the core power and cache power, and *total_phase_cycles* is the total number of cycles to execute a phase to completion. Table 2 shows some of the system's microarchitectural parameters that contribute to the EDP.

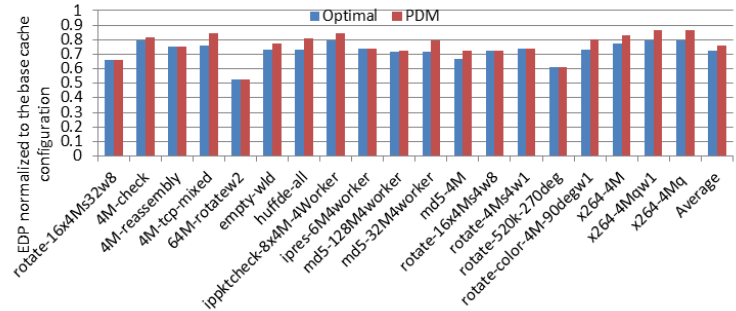

*(a)*

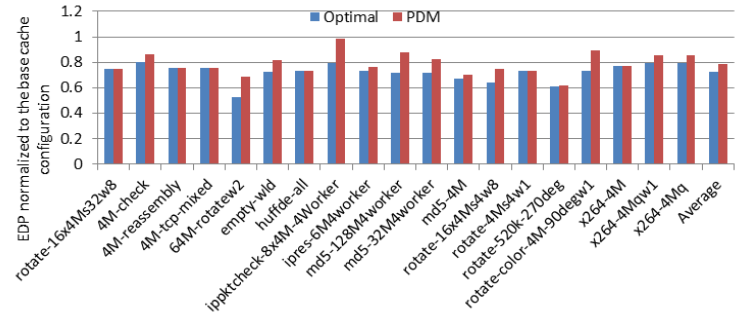

*(b)*

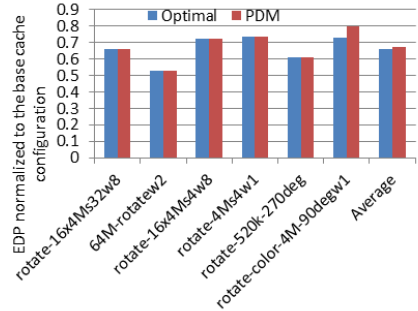

*(c)*

**Fig. 6:** EDP savings for the optimal configuration (Optimal) and the best configuration determined by phase distance mapping (PDM) normalized to the base configuration when using (a) *rotate-16x4Ms32w8* and (b) *huffde-all* as the base phase, and (c) when considering only the image processing phases. Phase distance mapping is also used to determine the configurations for the base phases, which shows the worst-case scenario for the base phases.

**5.2**   Results

5.2.1   PDM

Figure 6 (a) shows the EDP savings, as compared to the base configuration for the optimal configuration as determined using an exhaustive search (Optimal) and the best configuration as determined by phase distance mapping (PDM) for a single execution of each of the nineteen phases. *Rotate-16x4Ms32w8*, from the image processing domain, which rotates sixteen 4-megapixel greyscale images 90 degrees clockwise, is used as the base phase. On average over all phases, phase distance mapping achieved an EDP savings of 24%, with savings as high as 47% for *64M-rotatew2*, and was within 4% of the optimal configuration on average. PDM determined optimal configurations for seven of the nineteen phases, while some individual phases' configurations achieved EDP savings within 10% of the optimal.

To evaluate the effects that a different base phase has on the EDP savings, Figure 6 (b) shows the EDP savings, as compared to the base configuration, using *huffde-all* as the base phase. *Huffde-all* executes Huffman decoding on seven datasets. On average over all phases, phase distance mapping achieved an average EDP savings of 21%, with savings as high as 38% for *rotate-4Ms4w1*, and was within 7% of the optimal configuration. *Ippktcheck-8x4M-4Worker* had the lowest EDP savings (2%), as compared to the optimal (21%), because PDM selected a smaller line size than required. However, PDM still achieved some EDP savings over the base phase. Using *Huffde-all* instead of *rotate-16x4Ms32w8* as the base phase resulted in a 3% reduction in average EDP savings, while *64M-rotatew2*'s EDP savings dropped by 15%, and the number of phases for which PDM determined the optimal configurations reduced to five. The reduction in average EDP savings is due to the fact that *Huffde-all* is the only phase that performs any type of data compression whereas six of the phases perform image rotation. To verify this application-domain dependence when designating a base phase, we used *64M-rotatew2*, another image processing phase, as the base phase. For brevity, we omit the detailed results, but the results revealed that PDM using *64M-rotatew2* as the base phase achieved EDP savings that varied by less than 1% as compared to using *rotate-16x4Ms32w8* as the base phase.

We further analyzed the effectiveness of application-domain specialization by considering only the six image processing phases. Figure 6 (c) depicts the EDP savings normalized to the base configuration when considering only the image processing phases and using *rotate-16x4Ms32w8* as the base phase. The average EDP savings were 32%, which is 8% higher than the average over all nineteen phases, and were within 2% of the optimal, on average.

These analyses revealed that the magnitude of savings is highly application-domain dependent, and that even though good savings could be achieved by using any base phase, carefully considering the application domain when designating the base phase could significantly increase the EDP savings. Designating the base phase for a small, application-domain-specialized system with a small phase space can be done manually during design time, however, this manual designation is infeasible for large, general-purpose systems with a large phase space. For large systems, designers can use cluster analysis (e.g., k-means clustering [21] or graph-based models [43]) to partition the phase space into different domains, and the phase that most closely represents the largest cluster (most prominent domain) can be designated as the base phase.

5.2.2   DynaPDM

Figure 7 (a) shows the EDP savings of the optimal, PDM, and DynaPDM configurations normalized to the base cache configurations for a single execution of each workload/phase. To compare DynaPDM with PDM, we designated the base phase as *rotate-16x4Ms32w8*, which is from the image processing application domain. On average over all phases, DynaPDM achieved average EDP savings of 27% with savings as high as 47% for *64M-rotatew2*. DynaPDM determined the optimal configurations for 68% (thirteen) of the nineteen phases. On average over all phases, the EDP was within less than 1% of the optimal, with EDP savings within 3% of the optimal for *md5-4M*. DynaPDM showed a 4% improvement over PDM, however, PDM's savings are best-case savings acquired only after extensive design-time effort. Compared to PDM, DynaPDM increased the EDP for *rotate-16x4Ms4w8* and *rotate-4Ms4w1* by 1% and 2% respectively, because PDM leveraged domain specialization with a base phase from the image processing domain. However, DynaPDM improved EDP savings for eleven phases, with savings as high as 10% for *4M-tcp-mixed*.

To show DynaPDM's effectiveness in achieving significant EDP savings with any base phase, we quantified the EDP savings using *huffde-all* to represent an arbitrary base phase. We used *huffde-all* because *huffde-all* was the only phase that performed any form of code compression, and thus did not represent any of the other phases' domains. Figure 7 (b) depicts the EDP normalized to the base phase *Huffde-all*. Similar to using *rotate-16x4Ms32w8* as the base phase,

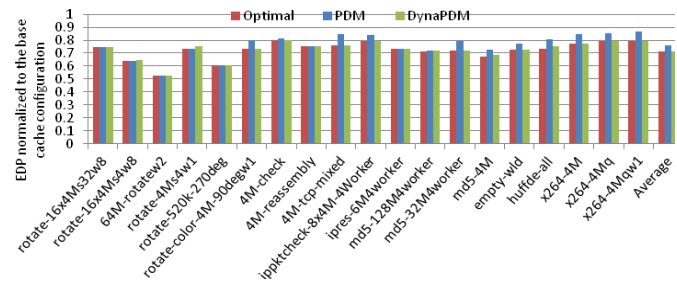

*(a)*

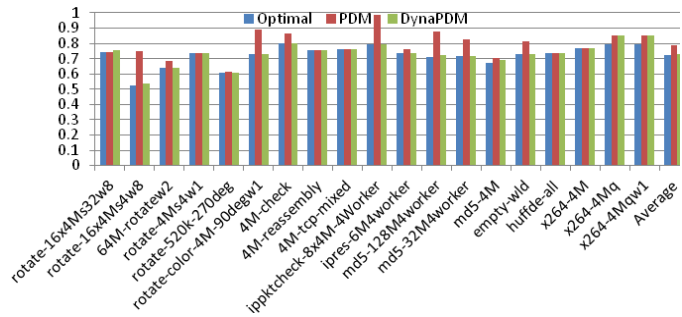

*(b)*

**Fig. 7** EDP savings normalized to the base configuration when using (a) rotate-16x4Ms32w8 as the base phase and (b) huffde-all as the base phase**.**

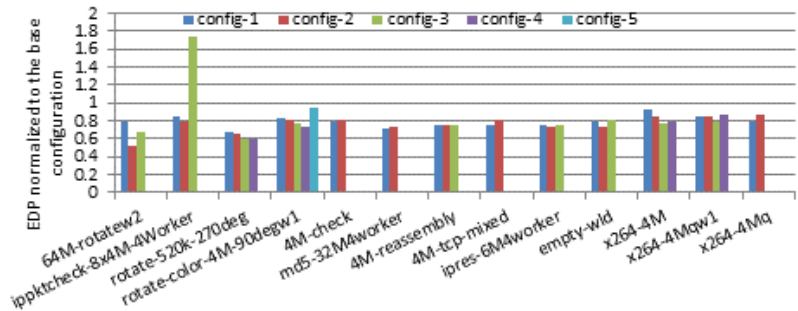

**Fig. 8** EDP savings of explored configurations normalized to the base configuration for all phases where DynaPDM determined the optimal configuration, where each bar, Config-1 to -5, represent the first to fifth configurations explored**.**

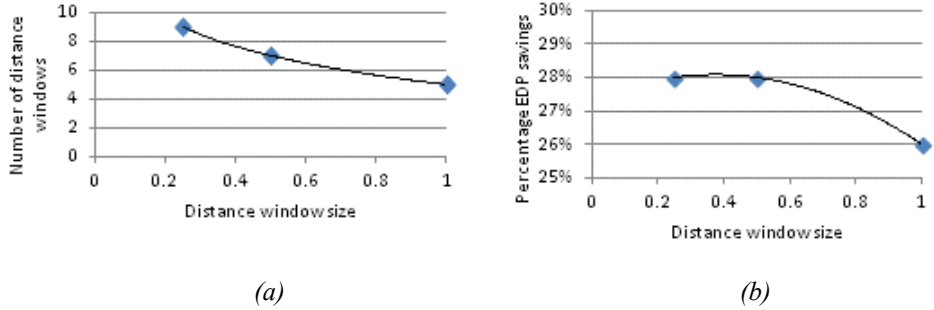

*(a)* *(b)*

**Fig. 9** Distance window size tradeoffs with (a) number of distance windows and (b) percentage EDP savings

DynaPDM achieved average EDP savings of 27%, with savings as high as 47% for *64M-rotatew2*. DynaPDM determined the optimal configurations for 63% (twelve) of the nineteen phases. Unlike PDM, where using a base phase that did not represent the prominent application domain adversely affected the EDP savings, DynaPDM's effectiveness in achieving EDP savings was independent of the base phase.

To show DynaPDM's ability to achieve significant EDP savings with minimal design space exploration and tuning overhead, we evaluated the configurations explored during tuning for all of the thirteen phases where DynaPDM determined the optimal configuration. Figure 8 shows the EDP savings for each of the explored configurations, Config-1 to -5, where each bar represents the single explored configuration's EDP normalized to the base configuration. Since inferior configurations are explored while finding the optimal configuration, the bars are not necessarily constantly decreasing. For *rotate-color-4M-90degw1,* DynaPDM explored only three configurations (less than 3% of the design space) before determining the optimal configuration, and then only explored two additional configurations. For twelve of the thirteen phases, all of the explored configurations reduced the EDP over the base configuration, preventing any tuning overhead, as compared to the base configuration, while determining those phases' optimal configurations. For example, for *rotate-color-4M-90degw1*, DynaPDM explored five configurations with EDP savings of 17%, 19%, 22%, 27%, and 4%, respectively, as compared to the base configuration. Since the fifth configuration reduced the EDP savings as compared to the first four configurations, DynaPDM determined the fourth configuration as *rotate-color-4M-90degw1*'s optimal configuration, since that configuration was the lowest EDP configuration. DynaPDM determined *rotate-color-4M-90degw1*'s optimal configuration without executing any configurations with higher EDP than the base configuration, thus minimizing tuning overhead. For *ippktcheck-8x4M-4Worker*, the third explored configuration reduced the EDP by 74% as compared to the base configuration,. However, since the second explored configuration achieved 21% EDP savings over the base configuration, DynaPDM determined that the second configuration explored was *ippktcheck-8x4M-4Worker'*s optimal configuration. In general, the results showed (details omitted for brevity) that DynaPDM only explored configurations that progressively increased the EDP savings for eighteen of the nineteen phases evaluated in our experiments, thus minimizing tuning overhead.

We also evaluated the impact of using a fixed versus a dynamic distance window size $S_d$. The distance window size $S_d$ determines the granularity/length of the distance windows and affects the distance window table's size (memory requirements), however, this size is minimal since only a few distance windows are created during the system's lifetime. Additionally, the distance window table's size can be fixed to adhere to system memory constraints, and a replacement policy, such as least recently used, can be used when the table is full. $S_d$ may also affect the configuration distances' accuracies (i.e., EDP savings). Larger $S_d$ reduces the number of distance windows and distance window table size, but may cause phases to map to distance windows that do not accurately represent the phases' characteristics, resulting in less accurate configuration distances.

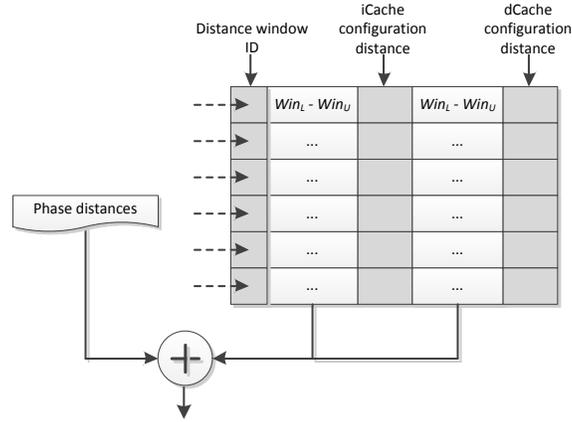

**Fig. 10** Distance window table architecture

**Table 3** Area and power consumption overhead for the distance window table

| # of entries | Total area (μm$^2$) | Area overhead | Total power (μW) | Power overhead |
|---|---|---|---|---|
| 16 | .517 | 0.24% | 12.78 | 0.11% |
| 32 | .620 | 0.30% | 14.18 | 0.11% |

Smaller $S_d$ increases the number of distance windows and the distance window table size, but may not necessarily increase EDP savings. Figure 9 (a) and (b) illustrate the tradeoffs of $S_d$ with the number of distance windows (i.e., distance window table size) and the percentage EDP savings, respectively. We empirically determined that $S_d = 0.5$ provided a good tradeoff between the number of distance windows and EDP savings. With $S_d = 0.5$, DynaPDM created seven distance windows, and achieved EDP savings within 1% of the optimal. With $S_d = 0.25$, DynaPDM created nine distance windows with no increase in EDP savings. Finally, with $S_d = 1$ DynaPDM created five distance windows, but the average EDP savings dropped to 26%. Therefore, using fixed distance window sizes offers no definitive advantage over using dynamic distance window sizes, while dynamic distance windows obviate the designer's need to specify a distance window size at design time.

## 6  Hardware Overhead

Figure 10 depicts the distance window table architecture, which comprises the only potential area/power overhead imposed by DynaPDM. The distance window table, which could be stored in the SRAM for quick access by the tuner, consists of the *n*-bit distance window identifications IDs, which identify unique distance windows for the instruction and data caches (iCache and dCache). *n* depends on the number of entries in the distance window table. For example, 16- and 32-entry distance window tables require 4-bit and 5-bit IDs, respectively. The distance window table also stores the phase distance ranges with minimum $Win_L$ and maximum $Win_U$ values, and associated configuration distances for the iCache and dCache. Since our configurable cache (Section 3) consists of eighteen different configurations, the distance window table only needs 5-bit entries, but could contain more entries for systems with additional configurations.

Since most embedded systems will typically only require a small number of entries in the distance window table (Section 4), we evaluated 16-entry and 32-entry distance window tables in synthesizable VHDL. We quantified the area and power consumption using Synopsis Design Compiler [39] and the Synopsis 90nm Generic Library. Table 3 depicts the area and power consumption values and overhead with respect to the MIPS32 M4K 90nm processor [25], which has an area of 0.21mm$^2$ and consumes 12mW of power at 200MHz. The 16-entry and 32-entry distance window tables result in negligible area and power overhead, imposing only a 0.24% and 0.30% area overhead, respectively, and a 0.11% power overhead. Given the negligible area and power overhead imposed by these tables, larger distance window tables will not attribute any appreciable area/power overhead.

## 7   Potential DynaPDM Usage scenarios

DynaPDM's ease of implementation and low power and area overheads makes DynaPDM especially suitable for embedded systems. One of DynaPDM's major usage scenarios is dynamic phase-based cache tuning. Based on our experience with using DynaPDM for cache tuning and DynaPDM's fundamentals as described in this work, we are optimistic that DynaPDM can be explored for other uses and areas of research. However, we note that DynaPDM's implementation in other optimization domains may require several additional considerations, depending on the specific usage scenario. In this section, we elaborate on other potential uses of DynaPDM in order to motivate future research directions.

To increase configurability and more closely satisfy design objectives, DynaPDM can tune other tunable hardware, such as clock frequency, instruction issue width, etc. Since different tunable hardware have different runtime behaviors, which may be susceptible to external factors, such as temperature, the specific hardware behaviors should be considered when adapting DynaPDM to these tunable hardware. To adapt DynaPDM for other tunable hardware, the application characteristics used to evaluate the phase distance must closely relate to the actually tunable hardware. For example, since clock frequency directly affects temperature, temperature can be used to evaluate the phase distance in dynamic voltage and frequency scaling (DVFS). Furthermore, since different tunable hardware have varying degrees of impact on different design objectives (e.g., clock frequency typically has a higher impact on IPC than cache miss rates), the specific design objectives must be considered simultaneously with DynaPDM's implementation and the tunable hardware's runtime behaviors.

For systems with several tunable hardware, DynaPDM can use multiple characteristics to evaluate the phase distance, wherein each characteristic can be weighted to reflect the impact of the tunable hardware on each characteristic. To incorporate multiple characteristics, the phase distance may be holistically calculated using a multidimensional distance calculation methodology, such as Euclidean distance, Manhattan distance, etc. with $N$ dimensions, where $N$ is the number of characteristics used to calculate the phase distance.

In heterogeneous core systems where energy efficiency and performance are affected by the specific application-to-core scheduling decisions, applications should be scheduled to cores that best satisfy design objectives. DynaPDM can be leveraged in these systems to determine the best application-to-core schedules based on the application characteristics. DynaPDM can evaluate the application distance (analogous to phase distance) between a previously executed application's characteristics and a new application's characteristics, and use the correlation between this application distance and core configurations to predict which core would achieve the best energy/performance efficiency for the new application.

Finally, in non-configurable systems where designers must determine the best configurations during design time, usually using extensive simulations and evaluations, DynaPDM can directly predict the systems' best configurations, which can then be compared to a base configuration to quantify the improvement (e.g., EDP and/or energy savings) over the base configuration. DynaPDM can also assist computer architecture researchers to determine optimal or near-optimal configurations without time consuming design-time simulations.

## 8   Conclusion and Future Work

Phase-based tuning specializes a system's configurations to varying runtime application characteristics to meet design constraints. One of the major challenges of phase-based tuning is determining the phases' best configurations without incurring significant tuning overhead. In this paper, we presented phase distance mapping—PDM—a phase-based tuning method that directly determines the best configuration for a phase with no design space exploration. Using extensive analysis of application phases and configurations, PDM determined configurations within 4% of the optimal configuration, with an average energy delay product (EDP) savings of 24%.

To reduce the design time overhead of pre-analyzing the application phases, and to make PDM more amenable to runtime changes and general purpose embedded systems with large or unknown phase spaces, we also presented a refinement to PDM, dynamic phase distance mapping—DynaPDM—a runtime phase-based tuning method that dynamically correlates a known phase's characteristics and best configuration with a new phase's characteristics to determine the new phase's best configuration, thereby reducing tuning overhead. DynaPDM adapts to runtime phase changes and eliminates designer effort. DynaPDM achieved average EDP savings of 27% and determined configurations within 1% of the optimal.

Future work includes evaluating DynaPDM's scalability to tunable many-core systems. In these systems, many cores could be tuned simultaneously, imposing large collective tuning overhead, and a single/centralized tuner could impose performance bottlenecks. Addressing these challenges may require using a dedicated tuner for each core (distributed tuning), which increases the area overhead, but may alleviate the performance bottleneck. Alternatively, the area overhead may be reduced by dividing the cores into clusters of cores comprising fewer cores, with a separate tuner for each cluster, which reduces the area overhead, but may impose performance bottlenecks. Thus, it is critical to evaluate the tradeoff between hardware overhead and shared resource contention. We also plan to explore and evaluate our proposed DynaPDM usage scenarios (Section 7) and more complex systems (e.g., heterogeneous systems, multilevel caches, etc.) with additional tunable hardware.

## Acknowledgment

This work was supported by the National Science Foundation (CNS-0953447). Any opinions, findings, and conclusions or recommendations expressed in this material are those of the author(s) and do not necessarily reflect the views of the National Science Foundation.